\documentclass[11pt]{article}
\usepackage{graphicx}
\usepackage{amsmath}
\usepackage{geometry}
\usepackage{natbib}
\usepackage[onehalfspacing]{setspace}
\usepackage{xcolor}
\usepackage{float}
\usepackage{hyperref}
\usepackage{enumitem}
\usepackage{booktabs}

\makeatletter

\begin{document}

\title{
Out-of-sample gravity predictions and trade policy counterfactuals\thanks{\setlength{\baselineskip}{1.2em}Acknowledgments: We are grateful to Achim Ahrens, Ramzi Chariag, José de Sousa, Miklós Koren, Ninon Moreau-Kastler, and György Ruzicska for helpful discussions and advice. We also thank participants in seminars at the LSE and Trier for many comments and suggestions. The usual disclaimer applies. Apfel, Breinlich, Santos Silva, and Zylkin gratefully acknowledge research support from the Economic and Social Research Council (ESRC grant ES/T013567/1). Novy gratefully acknowledges research support from the Economic and Social Research Council (ESRC grant ES/Z504701/1).}}\vspace{-0.25cm}

\author{Nicolas Apfel\thanks{University of Innsbruck; email: Nicolas.apfel@uibk.ac.at.}
\and
Holger Breinlich\thanks{University of Surrey; CEP, CESifo and CEPR; email: h.breinlich@surrey.ac.uk.}
\and
Nick Green\thanks{University of Surrey; email: nick.green@surrey.ac.uk.}
\and
Dennis Novy\thanks{University of Warwick; CAGE, CEP, CESifo and CEPR; email: d.novy@warwick.ac.uk.}\vspace{0.07cm} 
\and 
J.M.C. Santos Silva\thanks{University of Surrey; email: jmcss@surrey.ac.uk.} 
\and
Tom Zylkin\thanks{University of Richmond; email: tzylkin@richmond.edu.}
}

\date{\today\vspace{-0.5cm}}  

\maketitle

\begin{abstract}
\noindent Gravity equations are often used to evaluate the effects of trade policies, such as regional trade agreements. We argue that their suitability for this purpose critically depends on their ability to produce unbiased out-of-sample predictions. We propose a methodology to evaluate the out-of-sample predictions obtained with gravity equations and with machine learning methods. We find that the 3-way gravity model is difficult to beat when the purpose is to evaluate policy interventions, further cementing its position as the predominant tool for applied trade policy analysis. However, when the goal is to predict \emph{individual} flows, machine learning methods can be preferable.
\end{abstract}
\vspace{.5cm} 

\vspace{.25cm} 

\newpage

\section{Introduction}
\label{sec:introduction}

Gravity equations are the workhorse tool for evaluating the effects of trade policy interventions such as regional trade agreements (RTAs). In its basic form, the gravity model states that exports from country $i $ to country $j$ at time $t$ are a multiplicative function of exporter and importer GDPs and measures of bilateral distance. The effects of policy interventions such as RTAs are usually evaluated by including an additional dummy variable for whether the policy is in place at time $t$ for the pair $ij$. The coefficient on this variable then tells us by how much the policy increases trade, over and above what is predicted by the other variables.\footnote{Note that, when interpreted through the lens of a general equilibrium model, this coefficient estimate only provides a partial equilibrium effect.}  That is, the gravity equation is assumed to be the data generating process of the underlying counterfactual outcome in which the policy was not implemented.

A key justification for why gravity equations are well suited for providing this counterfactual is that they have very high explanatory power. However, such statements about the empirical performance of gravity equations usually relate to in-sample predictive performance, whereas out-of-sample predictive performance has not received much attention.\footnote{Recent exceptions include \cite{Verstyuk2022}, \citet{Koren_etal} and \citet{Kiyota_2025}. However, none of these contributions make the connection between out-of-sample performance and causal inference, which is the key focus of our paper. Another notable distinction is that we demonstrate some simple ways of retaining the information from all of the fixed effects estimates that feature in gravity estimations and verify their usefulness.} Of course, good in-sample predictive performance does not imply good out-of-sample performance if the former is achieved by overfitting statistical models.\footnote{Specifically, the danger of overfitting a model in-sample is that it approximates random noise in the data. Once we move to a different sample with a different realization of random noise, it will perform poorly.} 

We argue that to be able to provide reliable counterfactuals, gravity equations need to perform well out of sample, and we propose a data-based simulation methodology to evaluate the predictive performance of gravity equations and other forecasting methods. 

We then use our methodology to carry out a comparison between the out-of-sample forecasting performance of different specifications of the gravity equation and of a number of alternative statistical techniques from the machine learning literature.\footnote{Note that we will be using the terms predicting and forecasting interchangeably. Note also that we never consider prediction in the sense of forecasting trade data for periods for which we do not have estimates of the corresponding fixed effects.} These techniques provide a useful benchmark because they were designed with out-of-sample forecasting performance in mind. Thus, if the gravity equation does well against these alternative techniques, we can be confident that its current dominant position as the main tool for making forecasts and evaluating policy interventions is indeed merited.

Providing evidence on the relative out-of-sample predictive performance  of the gravity equation is important for two reasons. First, accurate trade flow predictions are directly useful for policymaking, for example when predicting the potential trade flows for a given pair of countries. This is illustrated by the `export potential assessment methodology' developed by the International Trade Centre in order to help countries identify promising products for export promotion activities (see \citealp{ITC_2025}). Secondly, and more importantly, we argue in Section \ref{sec:forecastcausality} that good out-of-sample predictive performance is crucial for obtaining reliable estimates of the causal impact of policy interventions. This is because the suitability of the gravity equation as a baseline counterfactual depends on its predictive performance out of sample, not its performance in sample. Put differently, if an alternative technique can provide better out-of-sample predictions for trade flows, then maybe this technique should take the place of gravity equations to generate counterfactual outcomes and evaluate the impact of policy interventions. In turn, this would call into question the reliability of a large literature on the trade effects of RTAs, currency unions, and WTO membership (see, e.g., \citealp{BaierBergstrand_2007}, \citealp{Rose_2001}, and \citealp{Rose_2004}).

Our results show that, when the goal is to obtain forecasts of individual trade flows, machine learning methods can significantly outperform standard gravity equations, especially in terms of mean squared prediction error. In contrast, we find that when the goal is to obtain estimates of the causal impact of trade policies, the 3-way gravity equation performs very well against machine learning methods, suggesting that it captures essential features of the data generating process of international trade flows. As for the other gravity models, our results show that they are generally outperformed by the machine learning methods. This underlines the importance of using a suitable model specification and the importance of the bilateral fixed effects. Given that the 3-way model is already considered to be the state-of-the-art version of the gravity equation (see \citealp{Yotov_etal_2016}, and \citealp{LSY}), we see our results as a further justification for its use. However, our results also suggest that there may be moderate gains from using a more flexible specification of the way trade costs and other pairwise characteristics enter the 3-way gravity model, and we demonstrate a simple, novel way of incorporating machine learning predictions within the 3-way gravity model in order to realize these gains.

The fact that the gravity equation can beat the machine learning methods when the goal is the estimation of policy effects, but not when the objective is to predict individual flows, is explained by the well-known bias-variance tradeoff: machine learning methods often trade increased bias for reduced variance through regularization. Estimates of the causal impact of trade policies average individual predictions, which reduces the relevance of the variance but not of the bias, thereby favoring the 3-way model whose predictions tend to have low bias. In the remainder of the paper, we will often emphasize this distinction between the use of predictions for estimation and for forecasting, which echoes the discussion in \cite{MullainathanSpiess2017}.

Our research contributes to several strands in the international trade literature. First, we contribute to the literature on using gravity equations as a tool to evaluate trade policy interventions (see \citealp{GoldbergPavcnik_2016}, and \citealp{Harrison_RC_2010}, for an overview). Our results show that the gravity equation does indeed merit its place as the dominant empirical tool in this area, albeit only in its most recent incarnation as the 3-way gravity model.

Second, we contribute to the literature on the estimation and interpretation of gravity equations (see, e.g., \citealp{Yotov_etal_2016}, and \citealp{LSY}). We contribute to this literature by pointing out that good out-of-sample forecast properties are an important but currently understudied criterion by which gravity equations should be evaluated. We also introduce some new methods for leveraging predictions from machine learning methods within the gravity framework and demonstrate their usefulness.

Finally, we also contribute to the smaller literature on the cross-sectional forecasting of international trade flows and other spatial flows. In contrast to much of this literature, our focus is on evaluating the out-of-sample predictive power of gravity equations, which we argue is more important than its in-sample predictive power. As such, our analysis complements the recent work of \citet{DingelTintelnotSEGS}, who study the predictive performance of spatial gravity models for predicting commuting flows.  In light of their finding that highly parameterized models can overfit relative to more parsimonious (covariates-based) specifications, it is notable that we find that the dense fixed effects used in 3-way gravity models are nonetheless useful for both prediction and treatment effect estimation, despite the large amount of variation they absorb.

The rest of this paper is structured as follows. Section \ref{sec:models} describes the forecasting methods that we will compare. Section \ref{sec:forecastcausality} establishes the conceptual link between 
out-of-sample forecasting performance and the reliability of trade policy causal impact estimates. Section \ref{sec:forecasts} introduces a general data-based simulation procedure for evaluating the forecasting performance of the different methods. 
Section \ref{sec:results} presents the results obtained when applying the proposed methodology. Section \ref{sec:conclusion} concludes.

\section{Statistical models for trade flows}
\label{sec:models}

We now describe the models for trade flows whose out-of-sample predictive performance we will compare.\footnote{Machine learning methods are predictive algorithms rather than parametric economic models. For simplicity, we often refer to both gravity equations and machine learning algorithms as ‘models’.} We start by presenting stylized versions of the gravity equations and then introduce the machine learning methods we consider.\footnote{In Subsection \ref{sec:data} we describe the additional regressors used in the models estimated in Section \ref{sec:results}.}

The first model we consider is the traditional gravity equation:
\begin{equation*}
y_{ijt}=\exp \left( \alpha _{0}+\beta _{1}\ln GDP_{it}+\beta _{2}\ln GDP_{jt} + %
\beta _{3}\ln dist_{ij}  +\tau D_{ijt}\right)+\varepsilon _{ijt},
\label{traditional}
\end{equation*}
where $y_{ijt}$ are exports from country $i$ to country $j$ at time $t$, $GDP_{it}$ and $GDP_{jt}$ denote the GDPs of the exporting and importing countries, $dist_{ij}$ is the distance between the two countries, $D_{ijt}$ is a dummy variable for the presence of treatment (e.g., an RTA dummy), and $\epsilon_{ijt}$ is the error term.

The second model we consider is the so-called 2-way gravity model which, following \citet{AvW_2001} and \cite{baldwinTaglioni}, augments the traditional gravity model by adding exporter-time and  importer-time fixed effects to account for multilateral resistance:
\begin{equation*}
y_{ijt}=\exp \left( \alpha _{it}+\gamma_{jt}+\beta _{3}\ln dist_{ij}+\tau D_{ijt}\right)+\varepsilon _{ijt}.  \label{twoway}
\end{equation*}
Note that the fixed effects absorb all country-specific determinants of trade flows, and the remaining covariates are conventionally interpreted as reflecting bilateral trade costs and other bilateral frictions that are specific to each pair.

The third model we consider is the 3-way gravity model which, as suggested by \citet{BaierBergstrand_2007}, augments the 2-way model by further including bilateral fixed effects:
\begin{equation*}
y_{ijt}=\exp \left( \alpha _{it}+\gamma _{jt}+\eta _{ij}+\tau D_{ijt}\right)+
\varepsilon _{ijt}.  \label{threeway}
\end{equation*}
The inclusion of bilateral fixed effects aims to mitigate the bias arising from possibly omitted time-invariant or slow-moving unobserved factors that influence both the probability of countries entering into RTAs and the amount of trade between them. 

In addition, we also consider a version of the gravity model which we call the 1-way model. This version includes bilateral fixed effects but omits exporter-time and importer-time fixed effects:

\begin{equation*}
y_{ijt}=\exp \left(\beta _{1}\ln GDP_{it}+\beta _{2}\ln GDP_{jt} +  \eta _{ij}+\tau D_{ijt}\right)+\varepsilon _{ijt}.  \label{oneway}
\end{equation*}
The purpose of considering this model is to evaluate the importance of controlling for the endogeneity of policy interventions, which both the 1-way and the 3-way models do, but the traditional and 2-way gravity models may not do. 

These gravity equations reflect conventional parameterizations for trade flows that have been widely used in the literature, and the 2-way and 3-way models are considered to be ``theory-based'' because of how they account for multilateral resistance (see \citealp{HeadMayer_2014}). We estimate the gravity equations using the Poisson pseudo maximum likelihood (PPML) estimator of \citet{Gourierouxetal1984}, as recommended by \citet{SantosSilvaTenreyro2006};  for models with fixed effects estimation is performed using the \texttt{ppmlhdfe} command of \citet{Correa_etal}.

While gravity equations provide a natural starting point, we also consider non-parametric machine learning methods. These methods are not grounded in models of international trade and, being non-parametric, do not permit the direct estimation of the effects of policy interventions. However, these methods are designed to have strong out-of-sample performance, which makes them particularly suitable to compute the imputation estimator discussed in Section \ref{sec:forecastcausality}.  

Applying machine learning estimators to our data requires choosing the set of regressors, or predictors, to be used. We consider two cases. We start by using just the covariates traditionally included in gravity equations and described in Subsection \ref{sec:data}  (e.g., GDPs and distance). Additionally, we also consider cases where we include as regressors the three sets of fixed effects estimated with the 3-way gravity model for the corresponding training sample, thereby ensuring that the machine learning methods have access to the same information as the 3-way specification. The advantage of including the estimated fixed effects is that they account for unobservable characteristics that are not easily controlled for otherwise. 

All the machine learning estimators were implemented using the user-written \texttt{pystacked} Stata command \citep{Ahrens_etal}, which in turn relies on Python’s \texttt{scikit-learn} library, see \cite{scikit-learn}. Specifically, we used the following methods.

 We begin by considering random forests. This technique consists of growing a large number of decision trees using bootstrap samples of the data and random subsets of predictors, and averaging their predictions (see \citealp{Hastie_etal}, chapter 15, for details). When we present our results in Section \ref{sec:results}, we refer to this approach as RF, or as RF-FE when the fixed effects are added to the set of predictors.

The second machine learning technique we use is gradient boosting, which sequentially fits shallow trees to the residuals from the current model, updating weights to minimize prediction error \citep[see][]{Friedman}. We again implement one version in which the estimated fixed effects from the 3-way gravity model are used as additional predictors (GB-FE) and one version without these regressors (GB).

Third, we estimate neural networks with and without using the estimated fixed effects from the 3-way gravity model as regressors; these are labeled NN-FE and NN, respectively. In contrast to the random forest and gradient boosting estimators, which are implemented using the default options in \texttt{pystacked}, for the neural networks estimators we standardize all predictors with the \texttt{StdScaler} pipeline because neural networks are sensitive to the scale of the regressors (see \citealp{LeCun_etal}), and use the Poisson loss because that option implies an exponential output activation function, avoiding negative fitted values.

The fourth machine learning technique we use is stacked generalization \citep{Wolpert92}. This is an ensemble method that combines the forecasts obtained using the three machine learning methods described earlier, by using them as explanatory variables in a non-linear least squares regression without an intercept, in which  the estimated coefficients are restricted to be non-negative and to sum to one (see \citeauthor{Breiman}, \citeyear{Breiman}, and \citeauthor{Hastie_etal}, \citeyear{Hastie_etal}). This, again, is the default option in \texttt{pystacked} but, to speed up the simulations, we only use four folds for the cross validation. As usual, we will present results using only the standard regressors, labeled SG, and results in which the fixed effects are added to the set of regressors, labeled SG-FE.

Finally, we estimate a ``3-way-ML'' gravity specification, which augments the 3-way gravity model by using as an added regressor the logarithm of the fitted values from the NN-FE estimator, which are necessarily positive. This specification allows us to assess whether the parametric 3-way model can be improved by incorporating flexible, nonlinear predictions from the machine learning methods.\footnote{%
Note that the predictions from the other machine learning methods are not necessarily positive, and therefore cannot be used to augment the 3-way model in the same way.} 
Because of the 3-way fixed effects, this augmented model allows these predictions to serve as an additional source of time variation in bilateral trade costs in addition to the explicit covariates.

\section{Out-of-sample forecast performance and causal estimates}
\label{sec:forecastcausality}

In this section we use the potential outcome framework of  \cite{Rubin1974} to establish the link between the out-of-sample forecast performance of gravity equations and their ability to estimate the causal effects of trade policy interventions.

Let the outcome of interest be $y_{ijt}$, the exports from $i$ to $j$ in period $t$, and let $D_{ijt}$ denote again the binary variable indicating the treatment status of pair $ij$ in period $t$. Moreover, define $y_{ijt}^{0}$ as the potential outcome without treatment and $y_{ijt}^{1}$ as the potential outcome with treatment. The observed exports can then be written as $y_{ijt}= y_{ijt}^{0}(1-D_{ijt}) + y_{ijt}^{1}D_{ijt}$.

As made clear by for example \cite{CianiFisher2019}, the parameter $\tau$ that is identified when a gravity equation is used to estimate the effect of $D_{ijt}$ is  closely related to the proportional treatment effect on the treated (PTT) defined as%
\begin{equation}
e^{\tau }=\frac{E\left( y_{ijt}^{1}|D_{ijt}=1\right) }{E\left(
y_{ijt}^{0}|D_{ijt}=1\right) },  \label{eq:ptt}
\end{equation}%
where $E\left(\cdot|D_{ijt}=1\right)$ denotes expectations taken over the population of treated units when they are treated (i.e, for which $D_{ijt}=1$). 

Given a suitable sample of size $N$, in which $n$ observations have $D_{ijt}=1$, the analogy principle (\citealp{goldberger68}, \citealp{Manski}) suggests that a consistent estimator of the PTT can be obtained by replacing the expected values in the numerator and the denominator of (\ref{eq:ptt}) with consistent estimators of these quantities.

Because for $D_{ijt}=1$ we have that $y_{ijt}=y_{ijt}^{1}$, the numerator of (\ref{eq:ptt}) is the expectation of outcomes that can be observed. Therefore, it can be estimated by the corresponding sample mean which, under very mild conditions, is an unbiased estimator of the quantity of interest, and is consistent when $n\rightarrow  \infty$. 

The difficulty lies in estimating the denominator of (\ref{eq:ptt}), which is the expectation of the unobserved counterfactual outcomes. To proceed, we need to select a method to predict $y_{ijt}^{0}$. Letting $\hat{y}_{ijt}^{0}$ denote the estimate of $y_{ijt}^{0}$ obtained by the chosen method, an estimator for the PTT can be constructed as 
\begin{equation}   
e^{\hat{\tau}}=\frac{n^{-1}\sum_{D=1}y_{ijt}}{ n ^{-1}\sum_{D=1}\hat{y}_{ijt}^{0}},  \label{estimator}
\end{equation}%
where the summations in both the numerator and denominator are across all $n$ observations with $D_{ijt}=1$, i.e., the \textquotedblleft treated\textquotedblright\ observations. If the denominator of (\ref{estimator}) consistently estimates $E\left(y_{ijt}^{0}|D_{ijt}=1\right)$ when $n\rightarrow  \infty$, expression (\ref{estimator}) gives a consistent estimate for the PTT.

It is clear that the consistency of $e^{\hat{\tau}}$ depends on the ability of the selected forecasting method to deliver unbiased predictions of $y_{ijt}^{0}$. Because $y_{ijt}^{0}$ is unobservable, the suitability of such a predictive method should be evaluated based on the properties of their out-of-sample predictions. This is because an estimator might have good predictive performance in sample but, due to overfitting, perform poorly when used to forecast quantities not observed in the sample, such as $y_{ijt}^{0}$. This clarifies why good out-of-sample forecast performance is important for obtaining consistent estimates of the causal effects of trade policy interventions. 

A sufficient condition for the predictions of $y_{ijt}^0$ to be unbiased is that they are generated by a method that correctly estimates its conditional expectation, which is the optimal predictor in a mean square error sense (see \citealp{Manski}). Therefore, machine learning methods designed to minimize the mean square prediction error can be seen as providing estimates of the conditional expectation of the data. However, because they are non-parametric and need to be able to handle high-dimensional data, these methods use regularization to increase precision, but this may introduce bias. In contrast, as argued by \cite{SantosSilvaTenreyro2006}, the gravity equation can be interpreted as a parametric specification of the conditional mean of trade flows. In this case, economic theory acts as form of regularization by directly providing a suitable conditional mean specification.

The current standard practice is to estimate the PTT using PPML to fit a gravity equation to the full sample, which leads to an estimate of $e^\tau$ where the denominator of (\ref{estimator}) is computed using the estimates obtained for the remaining parameters of the model. However, it has recently been pointed out that such estimates may not have a causal interpretation when the treatment effects are heterogeneous and not all pairs are treated at the same time (see, e.g., \citealp{Moreau-Kastler25}). This is because in this case the specification does not properly 
account for heterogeneous treatment effects,
implying a misspecification of the conditional expectation. To avoid this problem, in our simulations, we estimate all parametric and non-parametric models using only the observations with $D_{ijt} = 0$, and then use them to predict the counterfactual value of trade for the observations with $D_{ijt} = 1$. As discussed by \cite{Wooldridge2023}, \cite{Borusyaketal2024}, and \cite{Moreau-Kastler25}, under certain conditions, estimates obtained using imputation estimators of this kind can be given a causal interpretation.\footnote{As discussed above, the main condition is that  $n^{-1}\sum_{D=1}\hat{y}_{ijt}^{0}$ consistently estimates $E\left(y_{ijt}^{0}|D_{ijt}=1\right)$ when $n\rightarrow  \infty$. In turn, this condition is verified when there are proportional trends and no anticipation effects; see \cite{Moreau-Kastler25} for details.} 

While it is not the main focus of our paper, researchers might of course also be interested in accurate out-of-sample forecasts for other reasons. For example, they might want to predict the trade potential between two countries, as is the case for the ITC's export potential assessment tool previously mentioned (see \citealp{ITC_2025}). 

The accuracy of the out-of-sample predictions for $y_{ijt}^{0}$ is, therefore, important both when the goal is to estimate the PTT and when the goal is to forecast individual trade flows. However, there is an important difference between these two uses of out-of-sample predictions. If we are interested in forecasting \textit{individual} trade flows, it is desirable to have forecasts with the lowest possible mean squared prediction error, and therefore some bias is acceptable if it is accompanied by a significant enough increase in precision. In contrast, as seen from the denominator of (\ref{estimator}), if we are interested in obtaining consistent estimates of $e^\tau$, it is necessary that the out-of-sample predictions for $y_{ijt}^{0}$ are correct \textit{on average}, i.e., are unbiased. In this case, having noisier predictions is acceptable because what matters is the mean value of the predictions, whose variance vanishes as $n$ increases.

\section{Evaluating out-of-sample predictions}
\label{sec:forecasts}

In this section, we propose a general methodology for evaluating the predictive performance of different approaches. The key feature of this procedure is that it uses a data-based simulation design that is agnostic about the data-generating process to evaluate whether competing methods generate unbiased predictions of unobserved outcomes. As we have just seen, unbiased prediction of unobserved counterfactuals is the key requirement for consistency of the imputation estimator of the causal effect of policy interventions.

We start by introducing the dataset we use in our simulations, then describe the simulation procedure, and finally present the measures of predictive accuracy that we use.

\subsection{Data}
\label{sec:data}

We compare the out-of-sample performance of the competing models, described in Section \ref{sec:models}, on a standard and publicly available dataset of bilateral trade flows, that of \citet{yotov2025}.\footnote{These data can be downloaded from: \url{https://yotoyotov.com/Gravity_Undergrads.html}.} The dataset includes aggregate trade data, from the United Nation's Comtrade database, for the $100$ largest exporters in the world, covering $98.9\%$ of world exports, $97.7\%$ of world imports, and $98.3\%$ of world GDP between 1990-2023.  For our simulations, we only use data for 1994-2023 for pairs for which data are available every year. This gives us a balanced panel with $261,540$ observations out of the original $320,920$. The reason for using a balanced panel will become clear below.

Besides data on trade, the database contains standard explanatory variables that include a measure of distance, GDPs for importer and exporter, and indicators for membership of the European Union, membership of a customs union, regional trade agreements, contiguity, common language, and colonial ties, as well as an indicator for whether the importer has imposed a trade sanction on the exporter (see \citealp{yotov2025}, for more details on the data and their sources). All these variables are used in all regressions, although in some cases some of them drop out due to collinearity with the fixed effects.

\subsection{Simulation design} 
\label{sec:procedure}

To evaluate the out-of-sample predictions generated by each of the models, we proceed in four steps.

\begin{enumerate}

\item Split the $N$ observations in the dataset into a training sample of size $N-n$ (corresponding to the ``untreated'' observations for which $D_{ijt}=0$), and a test or hold-out sample of size $n$ (corresponding to the ``treated'' observations for which $D_{ijt}=1$). The way this crucial step is performed is described in detail below.

\item Use the $N-n$ ``untreated'' observations in the training sample to estimate the chosen model.

\item Using the estimated model from step 2, compute the out-of-sample predictions $\hat{y}_{ijt}^{0}$ for the $n$ ``treated'' observations in the test sample.

\item Compute measures of out-of-sample predictive performance for the estimators based on $K$ repetitions of steps 1 to 3; we discuss suitable measures of performance below. 

\end{enumerate}

We note that some of the machine learning methods we study use cross-validation-based regularization procedures, and this is done by minimizing the out-of-sample mean squared prediction error on a test dataset (also called a validation dataset). This is done by further splitting the $N-n$ observations in the training sample into secondary training and test samples. We stress that the original $n$ observations in the test sample chosen as described below are not available for this tuning procedure, ensuring that our prediction exercise is purely out of sample. 

The split of the data into the training and test samples is performed in three stages.

First, we define the probability with which pair $ij$ contributes some observations to the test sample. To construct this probability, we start by using the entire panel described above to estimate by PPML a 3-way gravity equation including all the available bilateral indicators. We then construct the probability that the pair $ij$ contributes observations to the test sample as
\begin{equation}\label{eq:probability}
      p_{i j}=1 /\left(1+\exp \left(\theta_1-\theta_2\hspace{0.03cm}\hat{\eta}_{i j}\right)\right),
\end{equation}
where $\hat{\eta}_{i j}$ are the estimated bilateral fixed effects from the 3-way regression, standardized to have mean zero and standard deviation one, $\theta_1$ is a parameter that controls the number of observations to be predicted, and $\theta_2$ is a parameter that controls the degree to which the probability of selection depends on time-invariant pair characteristics. Note that these probabilities are fixed over the $K$ repetitions of the procedure defined above and therefore this step is performed only once. 

Second, having computed $p_{i j}$, in each of the $K$ repetitions of the simulation procedure we select as actual contributors to the test sample the pairs for which $u_{ij} < p_{ij}$, where $u_{ij} \sim U(0,1)$ is drawn independently for each pair and each repetition. 

Third, and finally, for each pair with $u_{ij} < p_{ij}$, we randomly choose with equal probability a year between 2010 and 2014 to be the starting year of the test sample, which also includes all years after that. Because we are using a balanced panel that starts in 1994, this ensures that, for all pairs contributing to the test sample, we have between $16$ and $20$ observations in the training sample to estimate their pair fixed effects.

\subsection{Measures of predictive accuracy}
\label{sec:accuracy}

We consider a range of measures of predictive accuracy that allow us to compare different aspects of the out-of-sample predictive performance of the various methods described in Section \ref{sec:models}. We divide these into \textit{estimation} and \textit{forecasting} measures of accuracy, to distinguish between the use of out-of-sample predictions for estimation versus forecasting, as discussed in Section \ref{sec:forecastcausality}.

We start by considering \textit{estimation} measures of out-of-sample predictive accuracy, which are based on the value of the imputation estimator defined by (\ref{estimator}). Letting $\sum_{D_k=1}$ denote the sum over the set of $n_k$ observations in the test sample in repetition $k$ of the simulation procedure, the imputation estimator for the $k$-th repetition is computed as the ratio of the mean observed outcome to the mean predicted value for the observations not used for estimation, and takes the form
\begin{equation*}
\text{PTT}_k=\frac{n_k^{-1}\sum_{D_k=1}y_{ijt}}{n_k^{-1}\sum_{D_k=1}\hat{y}_{ijt}^0},
\end{equation*}%
where $\hat{y}_{ijt}^0$ represents the predicted value of $y_{ijt}$ estimated without using the $n_{k}$ observations excluded in repetition $k$. That is, as before, $\hat{y}_{ijt}^0$ is an estimate of $y_{ijt}^0$.

The first measure of out-of-sample forecasting accuracy that we consider is the mean of PTT$_k$ across the $K$ repetitions, denoted Mean\textsubscript{PTT}. Because the observations being predicted are not different from the ones used for estimation (i.e., because we perform a ``placebo'' exercise), good out-of-sample predictive performance is characterized by a value of Mean$_\text{PTT}$ close to $1$.\footnote{Our procedure can easily be modified to evaluate the effect of simulated treatments with different effects.} 

To gauge the variation of PTT$_k$, we also report its standard error across the $K$ simulation repetitions, denoted SE\textsubscript{PTT}, as well as its mean squared error, computed as
\begin{equation*}
\text{MSE}_{\text{PTT}}=\frac{1}{K}\sum_{k=1}^{K} (\text{PTT}_{k}-1)^2.
\end{equation*}%

As for the \textit{forecasting} measures based on the the ability of the model to predict individual observations, we consider the following three statistics.

\textbf{Mean Absolute Error} defined as 
\begin{equation*}
\text{MAE}=\frac{\sum_{k=1}^{K}\sum_{D_k=1}\left \vert \hat{y}_{ijt}^0-y_{ijt}\right \vert }{\sum_{k=1}^{K}n_{k}}.
\end{equation*}%

\textbf{Root Mean Squared Error} given by
\begin{equation*}
\mathrm{RMSE}= \sqrt{\frac{\sum_{k=1}^{K} \sum_{D_k=1}\left(\hat{y}_{ijt}^0-y_{i j t}\right)^2}{\sum_{k=1}^{K} n_{k}}}.
\end{equation*}

\textbf{Mean out-of-sample \textbf{\textit{R}}$^\mathbf{2}$} (Mean ${R^{2}}$) defined as the mean of the out-of-sample $R^{2}$ over the $K$ repetitions of the simulation procedure, where the out-of-sample $R^{2}$ is defined as the squared correlation between the observed and predicted values of trade over the $n_k$ observations in the test sample for repetition $k$ of the simulation process. 

Note that we report the MAE and the RMSE divided by the mean of the observed values in the test sample, $\left( \sum_{k=1}^{K} n_{k} \right) ^{-1} \sum_{k=1}^{K} \sum_{D_k=1} y_{i j t}$, so that their magnitudes are directly interpretable as percentages.

\section{Results}
\label{sec:results}

In this section, we present the results obtained by applying the procedure described above to compare the models presented in Section \ref{sec:models}. We consider two designs where the probabilities defined by (\ref{eq:probability}) are evaluated at different values of $\theta_1$ and $\theta_2$. 

\begin{enumerate}[label=(\roman*)]

\item In the first design, we set $\theta_1 = 5$ and $\theta_2 = 1$. We refer to this as the ``endogenous treatment'' case (see Table \ref{endog_case} below). It is endogenous in the sense that pairs with large $\hat{\eta}_{ij}$ are more likely to be selected, mirroring the fact that time-invariant pair characteristics are likely to play a role in the formation of RTAs, as suggested by \citet{BaierBergstrand_2007}.

\item In the second design, we set $\theta_2 = 0$ so that the probability of selection is independent of the bilateral fixed effects, and set $\theta_1 = 4.6$ so that the training and test samples have sizes comparable to those in case (i).  We refer to this as the ``exogenous treatment'' case (see Table \ref{random_case} below). While we believe that case (i) is the most empirically relevant setting, we  consider this case because it provides a useful benchmark and it can be realistic in some scenarios (e.g., when evaluating the impact of natural disasters).

\end{enumerate}

Table \ref{endog_case} shows results for the case of endogenous treatment (i.e., with $\theta_1=5$ and $\theta_2=1$) which, across the $1,000$ replications, yields values of $n_{k}$ ranging from $680$ to $1354$, with a mean of $1018.5$. 

Comparing the four gravity models first, we see that the 3-way model outperforms the other specifications in all criteria considered. Also of interest is the fact that the 2-way model performs worse than the 1-way model in terms of estimation. The most likely explanation for this is that the 2-way model does not control for the endogeneity of the treatment, because it does not include bilateral fixed effects. Interestingly, however, the traditional gravity model also outperforms the 2-way model even though it does not control for bilateral fixed effects. A plausible explanation for the latter finding is that not controlling for the multilateral resistance terms that motivate the country-time fixed effects used in the 2-way and 3-way models induces a downward bias that partially offsets the upward bias from not accounting for the endogeneity of the treatment.\footnote{Note that the results in Table 1 show that the estimates of the 2-way model are biased upward, whereas those from the 1-way model are biased downward.}

Moving on to the machine learning methods, we start by noting that the inclusion of the estimated fixed effects among the regressors dramatically improves the performance of these methods. This speaks to the importance of the information embedded in these fixed effect estimates. Looking at the estimation part of the table, we find that some of the machine learning methods can do as well, or even better, than the 3-way model in terms of Mean\textsubscript{PTT}, but have much higher SE\textsubscript{PTT}. In turn, the ensemble method augmented with the estimated fixed effects from the 3-way model, SG-FE, has a bias slightly larger than that of the 3-way model, but compensates for that by having a smaller SE\textsubscript{PTT}. Therefore, these two estimators have virtually identical MSE\textsubscript{PTT}.

\begin{table}[H]
\centering
\caption{Results for the case of endogenous treatment}
\label{endog_case}

\small
\resizebox{\textwidth}{!}{%
\begin{tabular}{lcccccccccccccc}
\toprule
\toprule
 & Trad & 2-way & 1-way & 3-way & SG & RF & GB & NN & SG-FE & RF-FE & GB-FE & NN-FE & 3-way-ML \\
\midrule
\multicolumn{14}{l}{\textbf{Estimation}} \\
\midrule
Mean\textsubscript{PTT}  & 1.028 & 1.154 & 0.964 & 0.993 & 1.007 & 1.007 & 1.069 & 1.039 & 1.031 & 1.042 & 1.007 & 1.005 & 1.000 \\
SE\textsubscript{PTT}      & 0.217 & 0.238 & 0.107 & 0.093 & 0.177 & 0.177 & 0.221 & 0.217 & 0.087 & 0.095 & 0.139 & 0.108 & 0.084 \\
MSE$_{\text{PTT}}\times 10$           & 0.480 & 0.805 & 0.127 & 0.086 & 0.314 & 0.314 & 0.538 & 0.485 & 0.085 & 0.107 & 0.192 & 0.117 & 0.070 \\
\midrule
\multicolumn{14}{l}{\textbf{Forecasting}} \\
\midrule
MAE         & 0.562 & 0.477 & 0.318 & 0.260 & 0.432 & 0.432 & 0.566 & 0.526 & 0.243 & 0.258 & 0.425 & 0.263 &0.244 \\
RMSE        & 2.360 & 2.310 & 1.840 & 1.559 & 2.348 & 2.348 & 2.448 & 2.245 & 1.026 & 1.217 & 1.554 & 1.280 & 1.114 \\
Mean $R^2$        & 0.718 & 0.781 & 0.890 & 0.914 & 0.791 & 0.791 & 0.718 & 0.747 & 0.913 & 0.906 & 0.817 & 0.913 & 0.917 \\
\bottomrule
\bottomrule
\end{tabular}
}

\vspace{-0.25cm}
\scalebox{.975}{
\begin{minipage}{1.00\textwidth}
\advance\leftskip 0cm
{\footnotesize \singlespacing {
\textit{Notes}: The ``Estimation'' part focuses on the goal of obtaining consistent estimates of the effects of policy interventions (average forecasts), whereas the ``Forecasting'' part focuses on the accuracy of forecasts for individual observations. The first four columns refer to gravity models (Traditional, 2-way, 1-way, 3-way). The remaining columns refer to machine-learning based methods. See the text for details.
} \par }
\end{minipage}
}
\end{table}

\newpage

The one area where the machine learning methods can outperform the 3-way model is in the forecasting of individual observations, which is seen in the forecasting part of the table. Here, the SG-FE approach comes out ahead of the 3-way model both for the MAE and RMSE, although not for the Mean $R^2$. However, the 3-way model outperforms almost all other machine learning methods in terms of the MAE, as well as in terms of Mean $R^2$.

Finally, we find that the 3-way-ML estimator outperforms all other estimators in terms of estimation criteria. In addition, it has the best Mean $R^2$ and generally performs well at prediction, notably performing better across all metrics than the NN-FE method that it draws from. Since this augmented model continues to include 3-way fixed effects, the improvement over the basic 3-way model can only come from the augmented model better capturing the effect of pairwise characteristics with time variation, which in the underlying trade model are usually interpreted as time-varying trade costs.\footnote{In our simulations, the gains of the 3-way-ML model with respect to the 3-way model are mainly on precision because the 3-way model is essentially unbiased in this setting. However, in generic empirical settings, it could be the case that the additional sources of trade costs captured by the 3-way-ML method are also correlated with the assignment of the treatment in ways not directly captured by the fixed effects and other covariates, in which case the 3-way-ML method could reduce bias as well.} These added components of trade costs can reflect interactions between standard trade cost covariates such as regional trade agreements and customs unions, but they can also include more complex interactions involving any of the variables used as predictors by the NN-FE method, including the fixed effects estimates from the original 3-way model as well as the GDPs of each country.
Therefore, our results suggest that when the aim is to evaluate policy interventions using an imputation estimator, there may be some benefit to utilizing machine learning methods, but this benefit turns out to only materialize when the predictions from these methods are embedded within the 3-way gravity framework so that the effects of trade costs and other pairwise characteristics can be modeled more flexibly.\footnote{Of course, there are other ways to model trade costs more flexibly within the gravity framework, as the underlying theory does not indicate a particular specification for trade costs. This can be done parametrically (e.g., by including interactions between standard gravity variables as regressors), or nonparametrically as we do in the 3-way-ML estimator. The advantage of the latter approach is that it allows for rich functional forms for trade costs without requiring the researcher to specify them parametrically, thereby avoiding the potential overfitting from highly parameterized specifications.}

Overall, these results illustrate the mean-variance tradeoff discussed before. Indeed, we see that machine learning estimators can provide predictions of individual observations with low RMSE. However, the averaging of these individual predictions to compute PTT$_k$ reduces the importance of the variance but not of the bias, leading estimators based on less biased predictions to dominate when the number of observations being averaged is sufficiently large.\footnote{We also consider a case with  $\theta_1=7.5$ and $\theta_2=1$, which leads to values of $n_k$ between $10$ and $203$, with a mean of $84.2$, and the results are qualitatively similar to those in Table \ref{endog_case}.}  

Table \ref{random_case} shows results for the case with exogenous treatment (i.e., with $\theta_1=4.6$ and $\theta_2=0$) which, across the $1,000$ replications, yields values of $n_{k}$ ranging from $718$ to $1,399$, with a mean of $1044.9$. The results from Table \ref{random_case} largely confirm those from Table \ref{endog_case}. The one notable difference to Table \ref{endog_case} is the better performance of the 2-way model, which is now comparable to the 3-way model in terms of the bias of the imputation estimator and outperforms both the traditional gravity model and the 1-way model in this respect. This confirms our earlier supposition that the endogeneity of the treatment explains the poor performance of the 2-way model in Table \ref{endog_case}. 

A natural interpretation of the simulation results reported in Tables \ref%
{endog_case} and \ref{random_case} is that flexible specifications of
theory-based gravity equations, possibly using machine learning predictions
as explanatory variables, can provide a good approximation to the
conditional expectation of trade flows. Because correctly specifying the
conditional expectation is a sufficient condition for the consistency of the
estimator, this makes the 3-way gravity model singularly well suited for
estimating the causal effects of trade policy interventions.

These findings are remarkable because nothing in the simulation design would 
\textit{a priori} suggest that the 3-way gravity model would produce
unbiased forecasts of trade flows. To illustrate why this is nontrivial,
consider if one were instead to model nonlinear transformations of trade
flows, such as their square root. In this case, the gravity equation would
no longer be expected to deliver unbiased forecasts because a model that
correctly specifies the conditional expectation of an outcome, in general,
does not also represent the conditional expectation of a nonlinear
transformation of that same outcome.\footnote{We confirmed this in unreported simulations; results available on request.}

\begin{table}[H]
\centering
\caption{Results for the case of exogenous treatment}
\label{random_case}

\small
\resizebox{\textwidth}{!}{%
\begin{tabular}{lcccccccccccccc}
\toprule
\toprule
 & Trad & 2-way & 1-way & 3-way & SG & RF & GB & NN & SG-FE & RF-FE & GB-FE & NN-FE & 3-way-ML \\
\midrule
\multicolumn{14}{l}{\textbf{Estimation}} \\
\midrule
Mean\textsubscript{PTT}  & 0.883 & 1.005 & 0.972 & 1.003 & 0.922 & 0.922 & 0.918 & 0.904 & 1.020 & 1.026 & 0.961 & 1.012 & 1.002 \\
SE\textsubscript{PTT}      & 0.234 & 0.242 & 0.140 & 0.115 & 0.210 & 0.210 & 0.251 & 0.236 & 0.113 & 0.120 & 0.180 & 0.133 & 0.111 \\
MSE$_\text{{PTT}}\times 10$          & 0.684 & 0.586 & 0.205 & 0.132 & 0.499 & 0.499 & 0.696 & 0.648 & 0.133 & 0.150 & 0.338 & 0.177 & 0.123 \\
\midrule
\multicolumn{14}{l}{\textbf{Forecasting}} \\
\midrule
MAE         & 0.639 & 0.519 & 0.330 & 0.268 & 0.491 & 0.491 & 0.670 & 0.602 & 0.255 & 0.270 & 0.477 & 0.274 & 0.259 \\
RMSE        & 3.348 & 3.299 & 2.565 & 2.158 & 3.435 & 3.435 & 3.480 & 3.221 & 1.460 & 1.711 & 2.152 & 1.756 & 1.742 \\
Mean $R^2$        & 0.701 & 0.782 & 0.882 & 0.906 & 0.773 & 0.773 & 0.690 & 0.720 & 0.903 & 0.896 & 0.795 & 0.904 & 0.908 \\
\bottomrule
\bottomrule
\end{tabular}
}

\vspace{-0.25cm}
\scalebox{.975}{
\begin{minipage}{1.00\textwidth}
\advance\leftskip 0cm
{\footnotesize \singlespacing {
\textit{Notes}: The ``Estimation'' part focuses on the goal of obtaining consistent estimates of the effects of policy interventions (average forecasts), whereas the ``Forecasting'' part focuses on the accuracy of forecasts for individual observations. The first four columns refer to gravity models (Traditional, 2-way, 1-way, 3-way). The remaining columns refer to machine-learning based methods. See the text for details.
} \par }
\end{minipage}
}

\end{table}

We thus conclude that if a researcher is interested in consistently estimating the causal effects of trade policy interventions such as RTAs, the 3-way gravity model is hard to beat as a workhorse framework. Any improvements in estimating these effects from using a machine learning-based method are likely to come from incorporating its predictions within the basic 3-way model rather than as a replacement for it.
It is only when the researcher is interested in predicting individual trade flows out of sample that machine learning techniques can improve upon the performance of the 3-way models. However, this improvement is only possible if these techniques incorporate the estimated fixed effects from a 3-way model as predictors; the machine learning models not making use of these fixed effects estimates do not offer such an improvement.\footnote{However, they often outperform the traditional gravity equation that uses the same information set.} Even then, if the ability to produce strictly positive predictions is preferred, the 3-way-ML method that embeds neural network predictions within the 3-way gravity model continues to be the best overall predictive method.

\section{Conclusion}
\label{sec:conclusion}

We argue that the ability of gravity equations to yield consistent estimates of the causal effects of policy interventions critically depends on their ability to produce unbiased out-of-sample predictions. We present a data-based simulation method to test this requirement, and we use it to evaluate the reliability of estimates obtained with parametric specifications of the gravity equation and with non-parametric machine learning forecasting methods.

We find that the 3-way gravity model produces essentially unbiased estimates of the counterfactual trade flows and therefore is difficult to beat when the purpose is to estimate the causal effects of trade policy interventions. Our results also suggest there may be moderate gains from using machine learning to model the trade costs part of the model more flexibly. 


The only area where machine learning methods can outperform the 3-way model is the prediction of individual trade flows, which is the purpose they are designed for. Given that the evaluation of policy interventions is arguably the most important reason for the use of the gravity equation in applied international trade research, we see our results as reassuring, in the sense that they further cement the role of the 3-way gravity model as the current state-of-the-art specification for gravity estimation. Future research will hopefully adopt our methodological approach and investigate whether our results hold in different datasets and using different machine learning estimators and possibly richer specifications of the gravity equation.

\bibliography{references}
\bibliographystyle{chicago}

\end{document}